\documentclass[twocolumn,floats,pra,aps,showpacs]{revtex4}
\usepackage{graphicx}
\usepackage{amsfonts}
\usepackage{amsmath}
\usepackage{amssymb}
\usepackage{exscale}
\usepackage{color}
\usepackage{subeqnarray}
\def\lan{\langle}
\def\ran{\rangle}
\def\va{\varepsilon}
\def\dag{\dagger}

\def\vk{{\bf k}}
\def\vK{{\bf K}}

\def\vp{{\bf p}}
\def\vq{{\bf q}}

\def\v0{{\bf 0}}
\newcommand{\bd}{\begin{equation}}
\newcommand{\ed}{\end{equation}}
\newcommand{\be}{\begin{equation}}
\newcommand{\ee}{\end{equation}}
\newcommand{\bt}{\begin{split}}
\newcommand{\et}{\end{split}}

\newcommand{\bn}{\begin{align}}
\newcommand{\en}{\end{align}}
\newcommand{\bea}{\begin{eqnarray}}
\newcommand{\eea}{\end{eqnarray}}
\newcommand{\ba}{\begin{array}}
\newcommand{\ea}{\end{array}}
\newcommand{\nn}{\nonumber}

\begin{document}

\title{Correlated Pair Approach to Composite Boson Scattering Lengths}
\author{Shiue-Yuan Shiau$^1$, Monique Combescot$^2$, and Yia-Chung Chang$^{3,1}$}\
\email[]{yiachang@gate.sinica.edu.tw}
\affiliation{(1) Department of Physics and National Center for Theoretical Sciences, National Cheng Kung University, Tainan, 701 Taiwan}
\affiliation{(2)Institut des NanoSciences de Paris, Universit\'e Pierre et Marie Curie,
CNRS, Tour 22, 2 Place Jussieu, 75005 Paris, France}
\affiliation{(3)Research Center for Applied Sciences, Academia Sinica, Taipei, 115 Taiwan}
\date{\today }


\begin{abstract}
We derive the scattering length of composite bosons (cobosons) within the framework of the composite boson many-body formalism that uses correlated-pair states as a basis, instead of free fermion states. The integral equation constructed from this physically relevant basis makes transparent the role of fermion exchange in the coboson-coboson effective scattering. Three potentials used for Cooper pairs, fermionic-atom dimers, and semiconductor excitons are considered. While the $s$-wave scattering length for the BCS-like potential is just equal to its Born value, the other two are substantially smaller. For fermionic-atom dimers and semiconductor excitons, our results, calculated within a restricted correlated-pair basis, are in good agreement with those obtained from  procedures numerically more demanding.  We also propose model coboson-coboson scatterings that are separable and thus easily workable, and that produce scattering lengths which match quantitatively well with the numerically-obtained values for all fermion mass ratios. These separable model scatterings can facilitate future works on many-body effects in coboson gases.

\end{abstract}


\maketitle

\section{Introduction}

The repeated interaction between two quantum particles is known to dress their interaction through an effective scattering commonly written in term of scattering length. Its determination is a major issue in the study of low-density bosonic gases. This study first focused on semiconductor excitons made of oppositely charged carriers\cite{Keldysh1968,HaugPRB1975,ShumwayPRB2001}. However, the achievement, in the 90's, of cold-atom Bose-Einstein condensates has triggered its study\cite{Pieri2000,Petrov2004,Brodsky2006,Birse2011,Alzetto2013} for dimers made of two fermionic atoms that differ by their hyperfine levels. More recently, the scattering of positronium atoms\cite{PlatzmannPRB1994,Ivanov2001,AdhikariPL2002,Chakraborty2004,Daily2015,Cassidy2005,Avetissian2014,Wang2014} and dark-matter particles\cite{Cline,Berezhiani} has attracted considerable attention due to their possible Bose-Einstein condensation. \

The major difficulty in coboson scattering problems is to properly include fermion exchange that occurs along with the repeated fermion-fermion interaction. Depending on the complexity of these fermion-fermion interactions, various approaches have been proposed: standard field theoretical or diagrammatic approach\cite{Keldysh1968,HaugPRB1975,Pieri2000,Brodsky2006,Alzetto2013}, quantum Monte Carlo method\cite{ShumwayPRB2001}, stochastic variational method\cite{Ivanov2001}, coupled-channel approach\cite{AdhikariPL2002,Chakraborty2004}, adiabatic hyperspherical method\cite{Daily2015}, renormalization group approach\cite{Birse2011}, and brute-force resolution of the four-body Schr\"{o}dinger equation\cite{Petrov2004}. All, except the coupled-channel approach, rely on a free-fermion formulation of the problem. This, of course, provides a very secure way to handle the Pauli exclusion principle. However, in doing so, one sacrifices the fact that the two-pair scattered state is very close to two single cobosons. To take advantage of this physical fact, one has to describe the four-body system in terms of cobosons while treating fermion exchange between cobosons in an exact way.\

In this work, we approach the four-body scattering problem through the coboson many-body formalism\cite{MoniqPhysreport,book} that was developed in the 2000's to address semiconductor excitons with Coulomb attraction between electrons and holes and equally strong Coulomb repulsion between electrons and between holes. We recently showed\cite{Moniq2015PRA} that this formalism takes a much simpler form when the interaction is restricted to attraction between different fermion species, as commonly used for fermionic-atom dimers, because the potential then reads as a one-body operator in the pair subspace (see Eq.~(\ref{eq:VintofBKp})). In that case, many-body effects are entirely driven by fermion exchange, as for Cooper pairs---a point not commonly understood.\

The coboson many-body formalism describes many-body systems in terms of correlated pairs. It is a natural approach to study Bose-Einstein condensation of composite bosons because, in the dilute limit, $N$ pairs in their ground state are very close to $N$ single ground-state pairs, within interaction terms depending on the (small) coboson density. $N$ correlated-pair states, however, have the unpleasant feature of forming an overcomplete basis. The coboson formalism circumvents this difficulty by using an operator algebra that manipulates cobosons via their creation operators.\

The great advantage of the coboson many-body formalism for scattering problems is threefold: first, it is conceptually simple; secondly, it renders physically transparent the subtle role played by fermion exchange between cobosons in the effective scattering; thirdly, it provides an easy way to numerically reach a good value of the coboson-coboson scattering length for arbitrary fermion mass ratios. To illustrate the power and flexibility of the method, we here use it to derive the coboson-coboson scattering length for two physically relevant potentials: (i) a short-range potential that acts between two different fermions having a finite total momentum, as for fermionic-atom dimers; (ii) the long-range Coulomb potential that acts between any charged fermions, as for semiconductor excitons. To grasp how the scattering length depends on the characteristics of the potential, we also consider a BCS-like short-range potential that acts between different fermionic atoms having a zero total momentum, similar to the one used for Cooper pairs. \

For the BCS-like potential, the scattering length can be analytically derived from the Richardson-Gaudin exact solution. The fact that this potential only allows zero-momentum pairs to interact restricts the interaction between cobosons to fermion exchange\cite{book}, and consequently forbids ladder-type processes. As a result, no dressing can occur by repeating the interaction and the scattering length reduces to its Born value.\

For the other two more complex potentials, we have derived the integral equation (\ref{laddereq}), from which the scattering length can be numerically obtained. The key kernel scattering $\zeta(_{mi}^{\, nj})$ (see Eq.~(\ref{eq:zeta})) contains the three fundamental scatterings, $\lambda(_{mi}^{\, nj})$, $\xi(_{mi}^{\, nj})$, and  $\xi^{in}(_{mi}^{\, nj})$ of the coboson many-body formalism which are visualized through the Shiva diagrams shown in Fig.~\ref{Fig1}(a,b,c). The $\lambda(_{mi}^{\, nj})$ and $\xi^{in}(_{mi}^{\, nj})$ scatterings involve a fermion exchange between the coboson fermionic constituents, while $\xi(_{mi}^{\, nj})$ is a direct interaction scattering, the $(m,i)$ cobosons being constructed on the same fermion pair. The effective scattering $\zeta(_{mi}^{\, nj})$ that appears in the kernel of the integral equation (\ref{laddereq}) clearly shows that, as for Cooper pairs, the interaction between fermionic-atom dimers is entirely governed by fermion exchange. In the case of semiconductor excitons, all three scatterings, ($\lambda,\xi,\xi^{in}$), are indispensable.

Restricting the intermediate correlated-pair relative-motion states in the integral equation (\ref{laddereq}) to the ground state renders the numerical resolution of this equation much simpler, while still providing good agreement with previous calculations of the scattering lengths done for fermionic-atom dimers and for semiconductor excitons in the positronium or hydrogen limit. In both cases, the scattering lengths as a function of fermion mass ratio exhibit a similar monotonously-increasing trend. Our results show that, as far as the scattering properties are concerned, the most important feature of the potential is not so much its short-range or long-range character as the number of internal degrees of freedom the interacting fermion pairs have, this number being one for zero-momentum pairs, and two for finite-momentum pairs.\

Furthermore, we propose a simple separable form for the kernel scattering, which provides an analytical expression for the scattering length. For both fermionic-atom dimers and semiconductor excitons, this analytical expression is in very good agreement with the full numerical results for  fermion mass ratio smaller than 10, a range corresponding to most physical cases. This simple and easily workable model scattering can be of great use to tackle complicated coboson many-body problems.

This paper is organized as follows: In Sec.~\ref{sec1}, we present the derivation of the integral equation for the coboson-coboson effective scattering in the framework of the composite boson many-body formalism, and its connection to the scattering length. In Sec.~\ref{sec2}, we present our results for the  scattering length in the case of the three potentials mentioned above. In Sec.~\ref{sec3}, we present the separable form we propose for the kernel scattering of the integral equation, and we compare the obtained results with the full solutions. Then, we conclude.

\section{Coboson-coboson scattering length\label{sec1}}

We consider a system made of two fermion species, $\alpha$ and $\beta$, which differ by their spins as for Cooper pairs, by their hyperfine indices as for fermionic-atom dimers, or by their band indices as for semiconductor excitons. The system Hamiltonian reads $H=H_0+V$. The kinetic part $H_0$ is  given by
\be
H_0=\sum_\vk \va_\vk^{(\alpha)} a^\dag_\vk a_\vk+\sum_\vk \va_\vk^{(\beta)} b^\dag_\vk b_\vk\, ,
\ee
 where ($a^\dag_\vk,b^\dag_\vk$) denote the ($\alpha,\beta$) fermion creation operators and $\va_\vk^{(\alpha,\beta)}=\hbar^2 \vk^2/2m_{\alpha,\beta}$ their energies.\

 The coboson many-body formalism\cite{MoniqPhysreport,book} is based on the single-pair eigenstates $|i\ran$ with $(H-E_i)|i\ran=0$. For translationally invariant systems, the coboson state index $i$ splits as $i=(\vK_i,\nu_i)$ where $\vK_i$  is the pair center-of-mass momentum and $\nu_i$ the pair relative-motion state index, the eigenstate energy being equal to $E_i=\hbar^2\vK^2_i/2M+\va_{\nu_i}$, with $M=m_\alpha+m_\beta$. The $i$-coboson creation operator reads $B^\dag_i=\sum_\vp B^\dag_{\vK_i\vp}\lan \vp|\nu_i\ran$, where $\lan \vp|\nu_i\ran$ is the pair relative-motion wave function, and $B^\dag_{\vK_i\vp}=a^\dag_{\vp+\gamma_\alpha \vK_i}b^\dag_{-\vp+\gamma_\beta \vK_i}$ with $\gamma_\alpha=1-\gamma_\beta= m_\alpha/(m_\alpha+m_\beta)$, creates a free-fermion pair with total momentum $\vK_i$ and relative-motion momentum $\vp$. \

Our procedure to obtain the coboson-coboson scattering length consists of two steps.\

\begin{figure}[t!]
\begin{center}
 \includegraphics[trim=2cm 4.2cm 0.5cm 4cm,clip,width=3.6in] {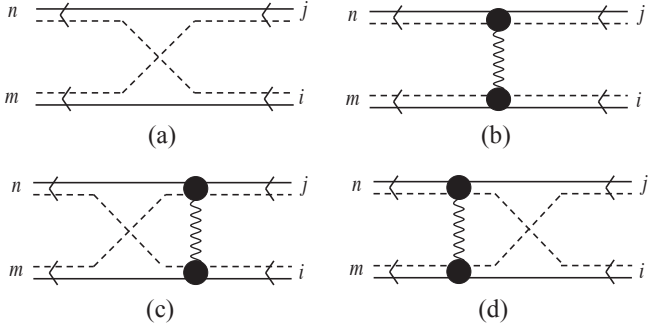}
   \caption{\small (a) Pauli scattering $\lambda(_{mi}^{\, nj})$ for fermion exchange in the absence of fermion-fermion interaction. (b) Direct-interaction scattering $\xi(_{mi}^{\, nj})$ in the absence of fermion exchange. (c) ``In" exchange-interaction scattering $\xi^{in}(_{mi}^{\, nj})$. (d) ``Out" exchange-interaction scattering $\xi^{out}(_{mi}^{\, nj})$. Solid lines represent $\alpha$ fermions, dashed lines represent $\beta$ fermions, and wavy lines represent interactions between the coboson fermionic constituents. }
   \label{Fig1}
\end{center}
\end{figure}

\noindent $(i)$ The first step is  to derive the ground-state energy of two fermion pairs by solving $(H-\mathcal{E}_2)|\Psi_2\ran=0$. To use the coboson many-body formalism, we first write $|\Psi_2\ran$ as $\sum_{ij} c_{ij} B^\dag_i B^\dag_j|v\ran$ with $c_{ij}=c_{ji}$ since $B^\dag_i B^\dag_j=B^\dag_j B^\dag_i$,  the vacuum state being denoted by $|v\ran$. The two commutators associated with fermion-fermion interaction are  
\bea
\big[H,B^\dag_i\big]_-&=&E_iB^{\dag}_{i}+ V^{\dag}_{i}\, ,\label{eq:HB}\\
\big[
V^{\dag}_{i},B^{\dag}_{j}\big]_-&=&\sum_{mn} B^{\dag}_{m}B^{\dag}_{n}\, \xi (_{mi}^{\, nj})\, ,\label{eq:VB}
\eea
where $\xi(_{mi}^{\, nj})$ is the direct-interaction scattering  shown in Fig.~\ref{Fig1}(b). This scattering consists of all possible fermion-fermion interactions between the fermionic constituents of the two cobosons, excluding interaction within each coboson.
Using the above two commutators in $(H-\mathcal{E}_2)|\Psi_2\ran=0$ yields
\be
0=\sum_{ij}\Big\{ c_{ij}(E_{ij}-\mathcal{E}_2)+\sum_{mn} \xi (_{im}^{ j\, n}) c_{mn} \Big\} B^\dag_i B^\dag_j|v\ran\label{ijcijeij}
\ee
with $E_{ij}=E_i+E_j$.  \

 To obtain a scalar equation for the $c_{ij}$'s, we project the above equation over $\lan v| B_pB_q$ using 
\be
\lan v|B_pB_qB^\dag_i B^\dag_j |v\ran
=\delta_{pi}\delta_{qj}-\lambda(_{p\, i}^{ q \, j})+(i\longleftrightarrow j)\, ,\label{eq:BBBB}
\ee
which is obtained from the other two commutators: 
\bea
\big[B_m,B^{\dag}_{j}\big]_-&=&\delta_{mi}-D_{mi}\, ,\\
\big[D_{mi},B^{\dag}_{j}\big]_-&=&\sum_nB^\dag_n \Big(
\lambda(_{mi}^{\, nj})+(i\longleftrightarrow j) \Big)\, .
\eea
The novel scattering of the coboson many-body formalism is the Pauli scattering $\lambda(_{mi}^{\, nj})$ associated with  fermion exchange induced by the Pauli exclusion principle between two cobosons. This scattering is quite unusual because it does not contain any fermion-fermion interaction; so, it is dimensionless.
As the topological symmetries of the Pauli scattering $\lambda(_{mi}^{\, nj})$ and the direct-interaction scattering $\xi (_{mi}^{\, nj})$ lead to  $\lambda (_{mi}^{\, nj}) =\lambda (_{nj}^{ m i})$ and $\xi (_{mi}^{\, nj}) =\xi (_{nj}^{ m i})$, Eq.~(\ref{ijcijeij}) ends up as
\be
0\!=\!(E_{pq}-\mathcal{E}_2)c_{pq}+\sum_{ij}\! \Big(\! \xi(_{p\, i}^{ q\, j})-\xi^{in}(_{p\, i}^{ q \, j})- (E_{ij}-\mathcal{E}_2)\lambda(_{p\, i}^{ q \, j}) \! \Big)c_{ij},\label{eq:cij}
\ee
 the exchange-interaction scattering being defined as $\xi^{in}(_{p\,i}^{ q\, j})
     =\sum_{mn} \lambda(_{pm}^{ q\, n}) \xi(_{m\,i}^{\, nj})$ (see Fig.~\ref{Fig1}(c)).\

 Since the two-pair ground state $|\Psi_2\ran$ is very close to two ground-state cobosons, $(B^\dag_0)^2|v\ran$, we are led to  write $\mathcal{E}_2$ as $2{\mathcal E}_1+\Delta$, where ${\mathcal E}_1$ is the single-pair ground-state energy. The $\Delta$ term we want to determine comes from the scattering between two cobosons, and so scales as the inverse of the sample volume $L^3$. Equation (\ref{eq:cij}) then reads
\be
0=(E_{pq}-E_{00}-\Delta)c_{pq}+\sum_{ij} \Big( \zeta(_{p\, i}^{ q\, j})+ \Delta\lambda(_{p\, i}^{ q \, j})  \Big)c_{ij}\, ,\label{eq:schrodingerDelta}
\ee
where the effective scattering  $\zeta(_{mi}^{\, nj})$ is given by
\begin{equation}
            \zeta(_{mi}^{\, nj})
     =\xi(_{mi}^{\, nj})-\xi^{in}(_{mi}^{\, nj})-(E_{ij}-E_{00})\lambda(_{mi}^{\, nj})\, .\label{eq:zeta}
       \end{equation}
This scattering contains the expected direct-interaction and exchange-interaction scatterings; it also contains a less obvious contribution constructed  on the dimensionless Pauli scattering $\lambda$ multiplied by an energy difference; so, in the case of semiconductor excitons, this energy part does not depend on the band gap, as physically reasonable. Also, note that this effective scattering has the required time-reversal symmetry, $\zeta(_{mi}^{\, nj})= \zeta^*(_{im}^{\,jn})$,
which follows from $\xi^{out}(_{mi}^{\, nj})=\big(\xi^{in}(_{im}^{ j\, n})\big)^*$ and the link between exchange-interaction scatterings\cite{MoniqPhysreport},
\begin{equation}
            \xi^{in}(_{mi}^{\, nj})+E_{ij}\lambda(_{mi}^{\, nj})= \xi^{out}(_{mi}^{\, nj})
     +E_{mn}\lambda(_{mi}^{\, nj})\, .\label{rel:xiinxiout}
  \end{equation}
Without the Pauli scattering $\lambda(_{mi}^{\, nj})$, the Hamiltonian constructed from this effective scattering would be non-hermitian, and thus unphysical (in the absence of external relaxation).\


To obtain $\Delta$ in Eq.~(\ref{eq:schrodingerDelta}), we separate the term in $(i,j)=(0,0)$ from the terms in $(i,j)\neq (0,0)$. For $(p,q) = (0,0)$, this readily gives, up to first order in $1/L^3$,
\be
\Delta c_{00}=\zeta(_{0\, 0}^{ 0\, 0})c_{00}+\sum_{ij\ne 0}\zeta(_{0\, i}^{ 0\, j}) c_{ij}\, .\label{eq:c00}
\ee
For $(p,q)\neq(0,0)$, the energy difference $E_{pq}-E_{00}$ for $\nu_p=\nu_q=\nu_0$ is equal to $\hbar^2(\vK^2_p+\vK^2_q)/2M$; so, it scales as $1/L^2$, while for $(\nu_p,\nu_q)\neq\nu_0$ this difference scales as $L^0$. As a result, since $\zeta/\lambda\sim 1$, all $\Delta$ terms in Eq.~(\ref{eq:schrodingerDelta}) are negligible because $\Delta$ scales as $1/L^3$.
 Dividing  Eq.~(\ref{eq:schrodingerDelta}) by $c_{00}$, with $c_{00}\neq 0$ since $(B^\dag_0)^2|v\ran$ constitutes the major part of $|\Psi_2\ran$, gives the integral equation fulfilled by $\hat{\zeta}(_{p\,0}^{q\,0})\equiv(E_{00}-E_{pq})c_{pq}/c_{00}$ as
\be
\hat{\zeta}(_{p\,0}^{q\,0})=\zeta(_{p\,0}^{q\,0})+\sum_{ij\neq00}\zeta(_{p\, i}^{q\, j})\frac{1}{E_{00}-E_{ij}}\hat{\zeta}(_{i\, 0}^{j\,0})\, . \label{laddereq}
\ee
Using Eq.~(\ref{eq:c00}), we end up with 
\be
\Delta= \zeta(_{0\, 0}^{ 0\, 0})+ \sum_{ij\ne 0} \zeta(_{0\, i}^{ 0\, j})\frac{1}{E_{00}-E_{ij}}\hat{\zeta}(_{i\, 0}^{j\,0})=\hat{\zeta}(_{0\,0}^{0\,0})\,, \label{app:delta}
\ee
which provides a contribution to the two-exciton energy from their interaction, up to first order in $1/L^{3}$.\

\noindent $(ii)$ In the second step, we associate the scattering length $a_s$ of two cobosons in a large volume $L^3$ with the two-pair ground-state energy $\mathcal{E}_2$ through 
\be
\Delta=\mathcal{E}_2-2\mathcal{E}_1= 4\pi \frac{\hbar^2 a_s}{M L^3}. \label{Delta}
\ee
To do it, we note that  Eq.~(\ref{eq:schrodingerDelta}) can be cast into a Lippmann-Schwinger equation
\be
c_{pq}=\delta_{p0}\delta_{q0}+G_{pq}\sum_{ij}\zeta(_{p\, i}^{ q\, j})c_{ij}\, ,
\ee
where $G_{pq}=1/(\mathcal{E}_2-E_{pq})$ is seen as the unperturbed Green function and $\sum_{ij}\zeta(_{p\, i}^{ q\, j})c_{ij}$ as the $T$-matrix element $T_{pq}$. Multiplying the above equation by $\zeta(_{m\, p}^{\, n\, q})$ and  summing over $(p,q)$ yield
\be
T_{mn}=\zeta(_{m\, 0}^{\, n\, 0})+\sum_{pq} \zeta(_{m\, p}^{\, n\, q})G_{pq}T_{pq}\, .
\ee
If, in $G_{pq}$, we replace $\mathcal{E}_2$ with $E_{00}$, we get $\Delta=T_{00}$ from Eqs.~(\ref{laddereq}) and (\ref{app:delta}). Drawing the relation between the $T$-matrix and the scattering length of two elementary particles\cite{Fetter}, we find the coboson-coboson scattering length $a_s$ as
\be
\Delta=\lim_{(pq)\rightarrow (00)}T_{pq}\equiv 4\pi\frac{\hbar^2 a_s}{ML^3}\, ,\label{delta_T_as}
\ee	
which proves Eq.~(\ref{Delta}).	\

Using Eqs.~(\ref{app:delta}) and (\ref{delta_T_as}), we end up with
\be
\hat{\zeta}(_{0\,0}^{0\,0})= 4\pi \frac{\hbar^2 a_s}{M L^3}\, .\label{Delta_as}
\ee
Equations (\ref{laddereq},\ref{Delta_as}) are the main results of this section. The scattering length in the Born approximation follows from $\zeta(_{0\,0}^{0\,0})$, while its value at all orders in interaction is obtained by solving Eq.~(\ref{laddereq}) exactly. This calculation requires the knowledge of the Pauli scattering $\lambda$ and the direct-interaction and exchange-interaction scatterings $\xi$ and $\xi^{in}$, all of which are constructed on single-pair eigenstates\cite{MoniqPhysreport}.


\section{Fermion-fermion potentials\label{sec2}}

Let us now calculate the scattering lengths induced by the three commonly-used potentials mentioned above.

\subsection{BCS-like potential}

We first consider the ``reduced BCS potential"
\be
V_{BCS}=-\sum_{\vp\vp'} B^\dag_{\textbf{0} \vp} v_{\vp-\vp'}B_{ \textbf{0}\vp'} ,\label{VBCS}
\ee
which acts between zero-momentum pairs, $B^\dag_{\textbf{0}\vp}=a^\dag_{\vp}b^\dag_{-\vp}$. The scattering is taken as short-ranged and separable, that is, $v_{\vp-\vp'}=v\, w_{\vp}w_{\vp'}$, with $w_\vp=1$ for $0\leq \va_\vp \leq \Omega $ and $w_\vp=0$ otherwise; $\va_\vp=\hbar^2 \vp^2/2\mu$, with $\mu^{-1}=m_\alpha^{-1}+m_\beta^{-1}$, is the pair relative-motion kinetic energy. (For Cooper pairs, , $w_\vp$ is taken equal to 1 for $\va_{F}-\Omega/2\leq \va_\vp \leq\va_{F}+ \Omega/2 $ and $w_\vp=0$ otherwise, with $\Omega\ll \va_{F}$, where $\Omega$ is of the order of a phonon energy and $\va_F$ is the normal-electron Fermi energy.)\

The single-pair ground-state energy $\mathcal{E}_1$ induced by this potential is known\cite{BCS,book} to follow, for a 3D density of states $\rho(\va)=\rho\sqrt{\va/\Omega}$, from
\bea
\frac{1}{v}&=&\sum_\vp\frac{w_\vp}{\va_\vp-\mathcal{E}_1}=\int^\Omega_0 \frac{\rho(\va)d\va}{\va-\mathcal{E}_1 }\nn\\
&=&2\rho\left(1-\sqrt{\frac{-\mathcal{E}_1}{\Omega}}\arctan\sqrt{\frac{\Omega}{-\mathcal{E}_1} }\right) \, .\label{sol_I1}
\eea
A bound-state solution, $ -\mathcal{E}_1>0$, exists provided that $v>v_{th}=1/2\rho$. The physically relevant regime corresponds to $|\mathcal{E}_1|\ll\Omega$, as obtained for $v$ close to $v_{th}$.  

The eigenstates of $H_0+V_{BCS}$ are also known for an arbitrary number of pairs\cite{rich,book}. The two-pair ground-state energy $\mathcal{E}_2$ is exactly given by $R_1+R_2$ with $R_1$ and $R_2$ solution of the two coupled Richardson-Gaudin equations
\be
\frac{1}{v}=\sum_\vp\frac{w_\vp}{\va_\vp-R_1}+\frac{2}{R_1-R_2}=(R_1\longleftrightarrow R_2)\, .\label{RGfor2pairs}
\ee
These equations are solved by rescaling $R_{j}$ as $\mathcal{E}_1(1-t_{j})$ with $j=(1,2)$ and by performing an expansion in $t_j$ using Eq.~(\ref{sol_I1}).
We end up with (see \ref{app:sec1})
\be
\Delta_{BCS}=-\mathcal{E}_1\left[4\pi\left(\frac{a}{L}\right)^3 +\mathcal{O}\left(\frac{a}{L}\right)^6\right] , \label{Delta_{BCS}}
\ee
 where $a$ is the single-pair ground-state Bohr radius, defined as $\mathcal{E}_1=-\hbar^2/2\mu a^2$. Equation (\ref{Delta}) then gives the corresponding coboson-coboson scattering length as
\be
a_s=\frac{ML^3 }{4\pi\hbar^2 }\Delta_{BCS}\simeq\frac{Ma}{2\mu}\, .\label{asBCSBorn}
\ee
This scattering length just corresponds to the Born value obtained\cite{Moniq2015PRA} for the $V_{CA}$ potential given in Eq.~(\ref{eq:VintofBKp}).\

We can understand why this is so by noting that zero-momentum pairs interact by Pauli blocking through fermion exchange, but not by fermion-fermion interaction\cite{book}. So, no dressing can occur by repeating the fermion-fermion interaction. As a result, BCS-like procedures such as Bogoliubov-de Gennes equations\cite{pieri2003PRL} or Ginzburg-Landau equations\cite{MeloPRL1993} can only produce the Born value of the coboson-coboson scattering length.\

\subsection{Cold-atom potential}

We now consider
\be
V_{CA}=
-\sum_{\vK\vp\vp'} B^\dag_{\vK\vp} v_{\vp-\vp'}B_{\vK \vp'}, \label{eq:VintofBKp}
\ee
where the $v_{\vp-\vp'}$ scattering, still equal to $v\, w_\vp w_{\vp'}$, now acts between fermion pairs having arbitrary center-of-mass momentum $\vK$. This potential reduces to $V_{BCS}$ for $\vK$ restricted to $\textbf{0}$. Since the single-pair ground state corresponds to $\vK=\textbf{0}$, the $\mathcal{E}_1$ energy obtained for $V_{CA}$ coincides with that for $V_{BCS}$.\

By contrast, the two-pair ground-state energy $\mathcal{E}_2$ differs from the one obtained for $V_{BCS}$ because $\vK\neq \textbf{0}$ states participate in the scattering between pairs. To get $\Delta_{CA}$, we numerically solve the integral equation given in Eq.~(\ref{laddereq}). Its first-order (Born) term $\zeta(_{0\,0}^{0\,0})$ was shown\cite{Moniq2015PRA} to read $-4\pi \mathcal{E}_1(a/L)^3$. The corresponding scattering length, $a_s^{(B)}=Ma/2\mu$, reduces to the well-known $2a$ value for equal fermion masses\cite{Haussman1993}. This Born value is substantially decreased by the repeated ladder-type processes, leading to the integral equation (\ref{laddereq}).

\begin{figure}[t!]
\begin{center}
\includegraphics[trim=0.5cm 0cm 1cm 0.5cm,clip,width=3.6in] {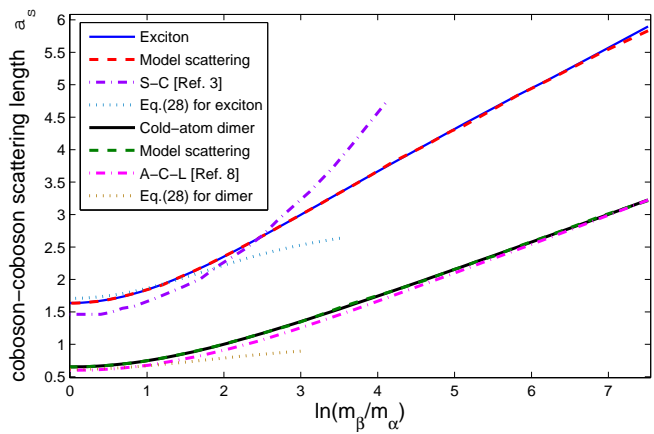}
   \caption{\small (color online) Coboson-coboson scattering length (in units of coboson ground-state Bohr radius $a$) for excitons and fermionic-atom dimers as a function of fermion mass ratio for $m_\beta /m_\alpha\geq 1$. }
   \label{Fig2}
\end{center}
\end{figure}
		
In the numerical resolution of the integral equation (\ref{laddereq}), we have restricted the relative-motion indices of the intermediate states $(i,j)$ to the ground state $\nu_0$. This approximation, which renders the numerical calculation far easier, actually catches the dominant processes, as shown below.  Momentum conservation in scattering processes then reduces the correlated-pair states $(i,j)$ to $(\vK',  \nu_0)$ and $(-\vK',  \nu_0)$; so, Eq. (\ref{laddereq}) becomes
\be
\hat{\zeta}(K,0)= \zeta(K,0)+\sum_{\vK'}\bar \zeta(K,K')G_0(K')\hat{\zeta}(K',0)\, ,\label{ladderK}
\ee
with $G_0(K)=(-2\hbar^2 K^2/2M)^{-1}$. As we here focus on $s$-wave scattering, we can average the kernel scattering over the $(\vK,\vK')$ angle, which amounts to replacing it with $\bar \zeta(K,K')=\frac 1 2 \int \sin\theta d\theta \zeta\big(_{\,(\vK,  \nu_0)\,\,\,\,\,\,\,(\vK',  \nu_0)}^{(-\vK,  \nu_0)\,(-\vK',  \nu_0)}\big)$. The resulting scattering length $a_s$, shown as a black solid curve in Fig.~\ref{Fig1}, yields $a_s\simeq 0.64a$ for equal fermion masses, instead of $\simeq 0.60a$ as obtained by previous procedures\cite{Petrov2004,Brodsky2006,Alzetto2013}, all of which are numerically far more demanding. By contrast, for large mass ratios, the $a_s$ curve, which essentially scales as $\ln (m_\beta/m_\alpha)$, coincides with that of Ref.~\onlinecite{Alzetto2013} (pink dash-dotted curve). The small discrepancy we find near equal fermion masses comes from the missing excited relative-motion states $\nu\neq \nu_0$. Their inclusion would significantly increase the numerical effort in solving Eq.~(\ref{ladderK}), for a precision unnecessary in view of the crude model potential that is used. We wish to mention that the present approach differs from the partial-bosonization procedure proposed in Ref.~\onlinecite{Pieri2000} which reported $a_s\simeq 0.75a$ for equal fermion masses, as can be seen from the different kernel scatterings $\zeta$  in the integral equations derived from the two procedures.


 \subsection{Coulomb potential}

 In the case of semiconductor excitons, the attractive part of the Coulomb potential between electrons  and holes  reads  
  \be
V_{eh}=-\sum_\vq v_\vq \sum_{\vk \vk'} a^\dag_{\vk+\vq}b^\dag_{\vk'-\vq}b_{\vk'} a_{\vk} ,
\ee
which is nothing but $V_{CA}$ except that $v_\vq$ now is equal to $4\pi e^2/(\epsilon_{sc}L^3q^2)$ for semiconductors having a dielectric constant $\epsilon_{sc}$. The long-range Coulomb potential also contains two similar repulsive parts between electrons and between holes. In this work, we consider that the two electrons (or the two holes) have the same spin; so, there is no restriction on fermion exchange between excitons.  \

The single-pair eigenstates correspond to those of  hydrogen atom. Their knowledge gives\cite{OBM} the Born term as $\zeta(_{0\,0}^{0\,0}) =-(26\pi/3)\mathcal{E}_1(a/L)^3$. The resulting scattering length $a^{(B)}_s=13M a/12\mu$, equal to $13a/3$ for equal carrier masses, has been first obtained by Keldysh and Kozlov\cite{Keldysh1968} using a different approach. The coboson formalism we here use allows us to go beyond the Born value.\

 The numerical resolution of the integral equation (\ref{laddereq}) is easy to perform by again restricting the intermediate relative-motion states to the ground state $\nu_0$. The results are shown in Fig.~\ref{Fig1} (blue solid curve). For equal carrier masses, we find $a_s\simeq 1.64a$ which agrees well with previous values ranging from $1.45a$ to $1.60a$ (see Table II of Ref.~\onlinecite{Daily2015}), as the inclusion of relative-motion excited states, $\nu\neq \nu_0$, is expected to reduce the scattering length by about $10\%$ [Ref.~\onlinecite{SYAnnals}]. Results for different carrier masses have only been reported by Shumway and Ceperley\cite{ShumwayPRB2001} (purple dash-dotted curve). The upward drift they found for large mass ratios was attributed to a systematic error. Here, we also find this drift, although not as dramatic as theirs. In the infinite hole mass limit, we find $a_s\simeq5.9a$; this result agrees well with a previous calculation\cite{SenEPL2006} performed in the ``static" approximation, that is, with $1s$ state only. Still, higher excitonic states $(2s,2p,\cdots)$ are known to weigh in more when holes become heavier and to reduce the scattering length, as necessary to recover the hydrogen-atom value, $\sim2a$ [Refs.~\onlinecite{SenEPL2006} and \onlinecite{Jamieson}]. Our results thus suggest a non-monotonous mass dependence, in contrast to fermionic-atom dimers. The study of this unexpected but interesting mass dependence is left for a future work.\

\begin{figure}[t!]
\begin{center}
\includegraphics[trim=0cm 0cm 0.5cm 0cm,clip,width=3.4in]{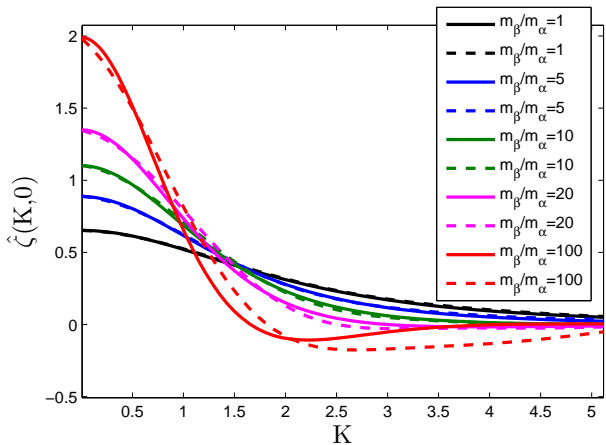}
   \caption{\small (color online) Renormalized fermionic-atom dimer-dimer scattering $\hat\zeta(K,0)$ (in units of $4\pi \hbar^2  a/ML^3$) as a function of the dimer center-of-mass momentum $K$ (in units of $1/a$). Various mass ratios are differentiated by colors. Solid and dashed curves correspond to exact, $\hat \zeta$, and model separable, $\hat\zeta^{(sep)}$, scatterings, respectively.}
   \label{Fig3}
\end{center}
\end{figure}
\begin{figure}[t!]
\begin{center}
\includegraphics[trim=0.3cm 0.2cm 0.5cm 0.2cm,clip,width=3.4in]{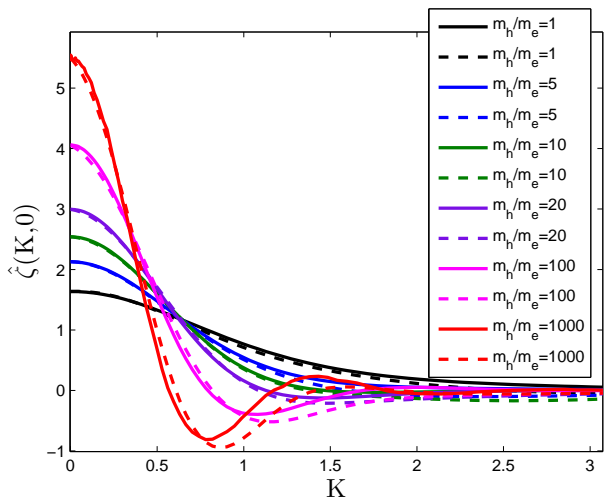}
   \caption{\small (color online) Same as in Fig.~\ref{Fig3} for the renormalized exciton-exciton scattering.}
   \label{Fig4}
\end{center}
\end{figure}	

 \section{Separable scattering model\label{sec3}}

 For practical use in future studies of coboson many-body effects, to have a kernel scattering $\bar \zeta(K,K')$ written in a separable form would be highly valuable. In particular, this would allow deriving an analytical expression for the scattering length and its mass dependence.\

 We can already obtain a very good agreement for $\bar\zeta(K,K')$ replaced by $\zeta_1(K)\zeta_1(K')$ with $ \zeta_1(K)=\zeta(K,0)/\sqrt{\zeta(0,0)}$, this separable scattering being exactly equal to $\bar \zeta(K,K')$ for $K$ or $K'$ equal to zero. The resulting scattering length then reads (see \ref{app:sec2})
 \be a_{s,1}^{(sep)}= \frac{ML^3\zeta(0,0)}{4\pi\hbar^2}\frac{1}{1+(M/\mu)\mathcal{F}}\, ,\label{as:sep1} \ee
where $\zeta(0,0)$ is equal to $-4\pi \mathcal{E}_1(a/L)^3$ for fermionic-atom dimers and to $-(26\pi/3)\mathcal{E}_1(a/L)^3$ for excitons. The factor $\mathcal{F} \equiv -(\mu/M)\sum_\vK G_0(K)\zeta^2_1(K)$, which results from repeated interactions, varies slowly with fermion mass ratio:  for fermionic-atom dimers, it goes from 0.503 when $m_\beta/m_\alpha=1000$ to 0.564 when $m_\beta/m_\alpha=1$, while for excitons, it stays essentially constant, going from 0.383 to 0.384. This simple analytic result is in fairly good agreement with the full result for $1\leq m_\beta/m_\alpha\leq 10$ (see Fig.~\ref{Fig1}), this ratio range covering most cases of physical interest.  \

With some more work, we can further improve the above results by considering the difference $\Delta_1(K,K')\equiv \bar\zeta(K,K')- \zeta_1(K) \zeta_1(K')$. This difference, which depends on mass ratio, consists of either a peak or a dip or both, spread along the $K=K'$ line (See Figs.~A\ref{fig2S} and A\ref{fig3S} in \ref{app:sec3}). The main peak can be modeled by a second separable scattering $\zeta_2(K) \zeta_2(K')$ with $ \zeta_2(K)=A_2 K e^{-\sigma_2^2(K-K_2)^2}$, while the remaining small shoulder/dip can be modeled by a third separable scattering $\pm  \zeta_3(K) \zeta_3(K')$ with $\zeta_3(K)=A_3 K e^{-\sigma_3^2(K- K_3)^2}$. The $(A_n, \sigma_n,K_n)$ parameters chosen to produce the best fit to $\Delta_1(K,K')$ are given in \ref{app:sec3}. We end with a renormalized scattering which reads as
\be
\hat{\zeta}^{(sep)}(K,0)= \zeta(K,0)+ \zeta_1(K)F_1 + \zeta_2(K)F_2 \pm  \zeta_3(K)F_3 \, ,  \label{sepa}
 \ee
the $F_n$'s being solution to three coupled equations (see \ref{app:sec2}). The resulting exciton-exciton and dimer-dimer scattering lengths shown in Fig.~\ref{Fig1} (red and green dashed curves) then are in excellent agreement with the full results over the entire mass-ratio range.

Figure \ref{Fig3} shows the numerically obtained renormalized fermionic-atom dimer-dimer scattering $\hat \zeta(K,0)$, as well as $\hat \zeta^{(sep)}(K,0)$ obtained from Eq.~(\ref{sepa}). Agreement is very good for all mass ratios, except for some minor discrepancy near $Ka\sim 2$ when $m_\beta/m_\alpha\ge 100$.

Figure \ref{Fig4} shows similar results for the exciton-exciton scattering. We again see very good agreement, except for some minor discrepancy near $Ka\sim 1$ when $m_h/m_e\ge 100$. The exciton-exciton  scattering behaves in a similar way as the dimer-dimer scattering shown in Fig.~\ref{Fig3}, although the characteristic $K$ range for dimers is twice as large (or twice as small in real space). This reflects a smaller dimer-dimer scattering length, being about half of the exciton-exciton value.


\section{Conclusion\label{sec4}}

The coboson many-body formalism, here used to obtain the ground-state energy of two fermion pairs, provides an easy way to numerically derive the coboson-coboson scattering length within a very good precision. The efficiency and flexibility of the procedure are demonstrated for short-range and long-range potentials. Another merit of this formalism is to reveal the physics entering the scattering processes that account for the scattering length, and to elucidate the precise role played by  fermion exchange in the effective coboson-coboson scattering. This approach can be extended to multiple-fermion particles, either bosonic or fermionic, such as atomic or semiconductor trions. In this work, we also propose a model scattering in a separable form that gives very good agreement with the numerically-obtained exciton-exciton and dimer-dimer scatterings for all fermion mass ratios. This simple and easily workable separable form should be very valuable for tackling many-body effects such as condensation of fermionic-atom dimers and semiconductor excitons.

\section*{Acknowledgement}

M.C. acknowledges many fruitful visits to Academia Sinica and NCKU, Taiwan. S.-Y.S. and Y.-C.C. have also benefited from various visits to INSP in Paris. Work supported in part by Ministry of Science and Technology, Taiwan under contract MOST 104-2112-M-001.

\renewcommand{\thesection}{\mbox{Appendix~\Roman{section}}} 
\setcounter{section}{0}

\renewcommand{\theequation}{\mbox{A.\arabic{equation}}} 
\setcounter{equation}{0} %
\section{Derivation of Eq.~(\ref{Delta_{BCS}})\label{app:sec1}}

We here show why two composite bosons that scatter via the $V_{BCS}$ potential given in Eq.~(\ref{VBCS}) have a scattering length which, in the large sample limit, is equal to the Born value $Ma/2\mu$ given by Eq.~(\ref{asBCSBorn}).\

 By rescaling $R_{j}$ as $\mathcal{E}_1(1-t_{j})$ with $j=(1,2)$, we are led to expand the Richardson-Gaudin equations given in Eq.~(\ref{RGfor2pairs}) as
\be
0=\sum_{n=1}^\infty t_1^n I_{n+1}+\frac{2}{t_1-t_2}=(t_1\longleftrightarrow t_2) \label{RGfor2}
\ee
with $I_n=\sum_\vp w_\vp \big[-\mathcal{E}_1/(\va_\vp-\mathcal{E}_1)\big]^{n}$. 
Through an integration by part, we find that the $I_n$'s obey the recursion relation
\be
0=(2n-3)I_n-2nI_{n+1}+2\rho\Omega\frac{(-\mathcal{E}_1/\Omega)^n}{(1-\mathcal{E}_1/\Omega)^n}\, .\label{appA:iterationIn}
\ee
 In the relevant regime $|\mathcal{E}_1|\ll \Omega$, this equation, taken for $n=1$, gives $I_2\simeq (-\mathcal{E}_1)(2\rho v-1)/2 v$, while for $n>2$, the last term of Eq.~(\ref{appA:iterationIn}) is negligible, and $I_n$ reduces to
\be
I_n\simeq \frac{(2n-5)!!}{2^{n-2}(n-1)!}I_2 \, .\label{appA:In}
\ee

To write $I_2$ in terms of $(L/a)^3$, we use (i) $\sqrt{-\mathcal{E}_1/\Omega}\simeq (2\rho v-1)/\pi\rho v$, as derived from Eq.~(\ref{sol_I1}); (ii) $L^3/2\pi^2=2\rho/\sqrt{\Omega}(2\mu)^{3/2}$, as obtained from $(L/2\pi)^34\pi\hbar^3 p^2dp=\rho \sqrt{\va_\vp/\Omega}d\va_\vp$ for $\va_\vp=\hbar^2\vp^2/2\mu$; and (iii) $\mathcal{E}_1=-\hbar^2/2\mu a^2$. Combining these results yields $I_2\simeq (L/a)^3/8\pi$. Equation (\ref{appA:In}) then gives, for $n>2,$
\be
I_n\simeq \frac{(2n-5)!!}{2^{n+1}(n-1)! \pi}\left(\frac{L}{a}\right)^3\,.\label{In_def}
\ee

To obtain $\mathcal{E}_2$ in terms of $(a/L)^3$, we sum and subtract the two Richardson-Gaudin equations given in Eq.~(\ref{RGfor2}). We get
\bea
0&=& \sum_{n=1}^\infty (t_1^n+t_2^n) I_{n+1}, \, \label{appRG1}\\
-4&=&(t_1-t_2)\sum_{n=1}^\infty (t_1^n-t_2^n) I_{n+1}\, .\label{appRG2}
\eea
Since  $(t_1,t_2)$ are expected to scale as $(a/L)^3$, we can truncate the sums up to their quadratic terms as
\bea
0&\approx & 4(t_1+t_2)+(t_1^2+t_2^2)\, ,\\
-32\pi\left(\frac{a}{L}\right)^3&\approx & (t_1-t_2)^2\, .\label{t1t2_1}
\eea
By combining these two equations, we get, for $t=t_1+t_2$,
\be
0=t^2+8t-32\pi\left(\frac{a}{L}\right)^3\, .
\ee
Its physical positive solution reads $t=4\pi (a/L)^3+\mathcal{O}(1/L^6)$. This gives the energy difference $\mathcal{E}_2-2\mathcal{E}_1 $ as $\Delta_{BCS}=-\mathcal{E}_1(t_1+t_2)=2\pi \hbar^2 a/\mu L^3 +\mathcal{O}(1/L^6)$. Using Eq.~(\ref{Delta}), the scattering length $a_s$ then reads $Ma/2\mu$.\

 Let us now consider cubic terms in $t_1$ and $t_2$ to see if they affect the above leading term. Equations (\ref{appRG1}) and (\ref{appRG2}) then read
\bea
0 \!&=&\!  8(t_1+t_2)+2(t_1^2+t_2^2)+(t_1^3+t_2^3)\,,\hspace{1cm} \label{t_1t_2_2}\\
-128\pi\left(\frac{a}{L}\right)^3\!&=& \!4(t_1-t_2)^2+ (t_1-t_2)^2(t_1+t_2)\, .\label{t1t2_3}
\eea
We first rewrite Eq.~(\ref{t_1t_2_2}) in terms of $t_1+t_2=t$ and then use
Eq.~(\ref{t1t2_3}) for $(t_1-t_2)^2$. This gives
\be
512\pi\left(\frac{a}{L}\right)^3=128\left(1+3\pi\left(\frac{a}{L}\right)^3\right)t+48t^2+8t^3+t^4\, .
\ee
The above equation only has one real positive solution $t=4\pi (a/L)^3+\mathcal{O}(1/L^6)$. This leads us to conclude that no ladder-type dressing occurs for the $V_{BCS}$ potential: the scattering length stays equal to the Born value obtained  for the $V_{CA}$ potential given in Eq.~(\ref{eq:VintofBKp}).
	
\renewcommand{\figurename}{Figure A\!}
\setcounter{figure}{0} %
\section{Modeling the exciton-exciton and dimer-dimer scatterings through a separable form\label{app:sec2}}

The angular-averaged scattering $\bar \zeta(K,K')$ in Eq.~(\ref{ladderK}) can be well represented by a sum of three separable scatterings as
\be
\bar \zeta^{(sep)}(K,K'){=} \zeta_1(K)\zeta_1(K'){+} \zeta_2(K) \zeta_2(K'){\pm}  \zeta_3(K)\zeta_3(K') \, ,
\ee
where  $\zeta_1(K)=\zeta(K,0)/\sqrt{\zeta(0,0)}$ and $\zeta_n(K)=A_n K e^{-\sigma_n^2(K- K_n)^2}$ for $n=(2,3)$. Details on how to choose the best-fit parameters $(A_n,\sigma_n,K_n)$ are given in \ref{app:sec3}. The resulting renormalized exciton-exciton scattering then reads as in Eq.~(\ref{sepa}), with $F_n\equiv\sum_{\vK} \zeta_n(K)G_0(K)\hat \zeta^{(sep)}(K,0)$. To obtain  $F_n$'s, we multiply Eq.~(\ref{sepa}) by $G_0(K)\zeta_n(K)$ with $n=(1,2,3)$, and we sum over $\vK$. This leads to three coupled linear equations
\be
\left[   \begin{array}{lll} 1- G_{11} & -G_{12} &  \mp G_{13} \\  -G_{21} & 1-G_{22} & \mp G_{23} \\  -G_{31} & -G_{32} & 1 \mp G_{33} \end{array} \right] \left[ \!\begin{array}{l} F_1 \\ F_2 \\ F_3 \end{array}\! \right] {=}\sqrt{\zeta(0,0)}  \left[\! \begin{array}{l} G_{11} \\ G_{21} \\ G_{31} \end{array}\! \right] \label{eq:Fimatrix}
\ee
with $G_{nm}=\sum_{\vK}\zeta_n(K)G_0(K)\zeta_m(K)$. Equation (\ref{sepa}) gives, since $\zeta_1(0)=\sqrt{\zeta(0,0)}$ while $\zeta_2(0)=0=\zeta_3(0)$,
\be
\hat \zeta^{(sep)}(0,0)=\zeta(0,0)\Big(1+F_1/\sqrt{\zeta(0,0)}\Big)\, .\label{app:zetasep}
\ee
By using Eq.~(\ref{Delta_as}), we get the scattering length $a_s^{(sep)}$ as $(ML^3/4\pi)\hat \zeta^{(sep)}(0,0)$.
\

 If we just keep the leading separable scattering, only $G_{11}$ survives in Eq.~(\ref{eq:Fimatrix}), and we find $F_1=\sqrt{\zeta(0,0)} G_{11}/[1-G_{11}]$. Inserting this $F_1$ value into Eq.~(\ref{app:zetasep}), we get $\hat \zeta_1^{(sep)}(0,0)=\zeta(0,0)/(1-G_{11})$. The corresponding scattering length $a_{s,1}^{(sep)}$ is given in Eq.~(\ref{as:sep1}).

\renewcommand{\theequation}{\mbox{C.\arabic{equation}}} 
\setcounter{equation}{0} %
\section{Comparison between the separable model scattering and the true scattering\label{app:sec3}}

\begin{figure}[t!]
\begin{center}
 \includegraphics[trim=4cm 1.5cm 4cm 15cm,clip,width=3.4in] {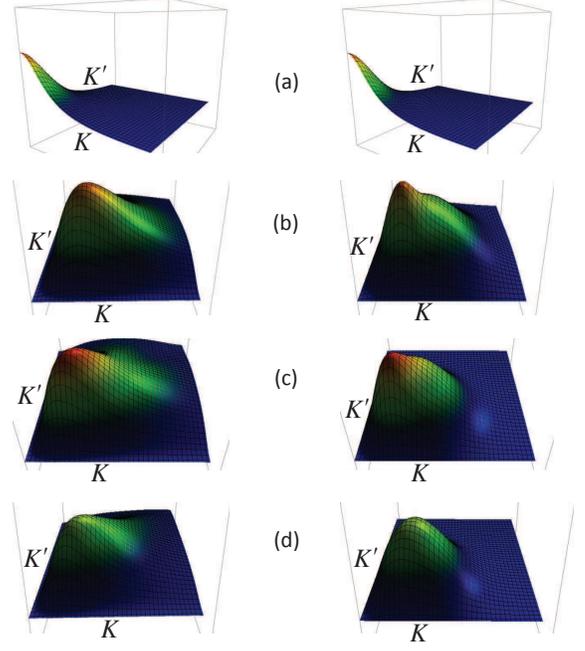}
   \caption{\small (color online) (a) True dimer-dimer scattering (left) and the leading separable scattering $ \zeta_1(K)\zeta_1(K')$ (right) for $m_\beta/m_\alpha=1$. (b) Difference $\Delta_1(K,K')$ (left) and the second separable scattering plus the third $ \zeta_2(K)\zeta_2(K')+ \zeta_3(K) \zeta_3(K')$ (right), for $m_\beta/m_\alpha=1$. (c)  Same as (b) for $m_\beta/m_\alpha=10$. (d)   Same as (b) for $m_\beta/m_\alpha=100$. The range of $K$ and $K'$ is $(0,10a^{-1})$.}
   \label{fig2S}
\end{center}
\end{figure}
\subsection{Dimers}

Figure A\ref{fig2S} shows the true angular-averaged scattering $\bar \zeta(K,K')$ and the leading separable scattering $ \zeta_1(K)\zeta_1(K')$ with $\zeta_1(K)=\zeta(K,0)/\sqrt{\zeta(0,0})$ for fermionic-atom dimers as well as their difference $\Delta_1(K,K')$. As seen in Fig.~A\ref{fig2S}(a), the leading separable scattering already looks fairly close to the full scattering, the difference being one order of magnitude smaller. \

We can improve the result by simulating the difference $\Delta_1(K,K')$, which contains a single peak stretched along the line $K=K'$. The peak can be approximated by a sum of two separable scatterings $\zeta_2(K)\zeta_2(K')+\zeta_3(K)\zeta_3(K')$ where $\zeta_n(K)=A_n K e^{-\sigma_n^2(K-K_n)^2}$ for  $n=(2,3)$.
We obtained a good fit with $\sigma_2=0.7a$ for all fermion mass ratios and $K_2 =(K_p^2-1/2\sigma_2^2)/K_p$, where $K_p$ is related to the position $K_m$ of the $\Delta_1(K,K')$ maximum through $K_p=K_m/\eta$. The $\eta$ factor is equal to $1.32$ for $m_\beta/m_\alpha<20$, and linearly increases as $0.2\ln(m_\beta/m_\alpha)$ for larger mass ratios.

A third separable scattering has been added to simulate the difference $\Delta_2(K,K')\equiv \Delta_1(K,K')-\zeta_2(K)\zeta_2(K')$ with $K_3=(K_m^2-1/2\sigma_3^2)/K_m$, where $K_m$ is the position of the $\Delta_2(K,K')$ maximum. $\sigma_3$ is equal to $0.2a$ for $m_\beta/m_\alpha=1$, then linearly increases as $0.02a\times (m_\beta/m_\alpha)$ until it reaches $0.3a$ at $m_\beta/m_\alpha=5$, and then it remains constant for all larger mass ratios. The amplitudes $A_2$ and $A_3$ are adjusted to match the maximum value of $\Delta_1(K,K')$ and minimize the difference in the small $K$ regime (see Fig.~A\ref{fig2S}(b,c,d)). \

\begin{figure}[t!]
\begin{center}
\includegraphics[trim=4cm 1.7cm 4cm 14.5cm,clip,width=3.4in] {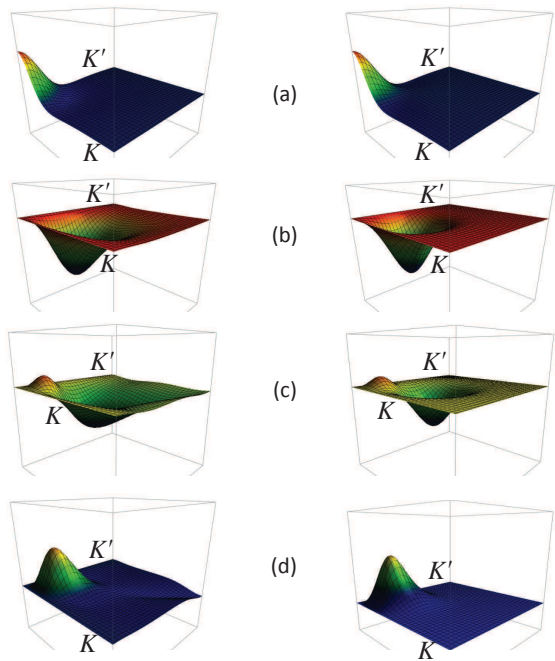}
   \caption{\small (color online) (a) True exciton-exciton scattering (left) and the leading separable scattering $ \zeta_1(K)\zeta_1(K')$ (right) for $m_h/m_e=1$. (b) Difference $\Delta_1(K,K')$ (left) and the third separable scattering $-\zeta_3(K) \zeta_3(K')$ (right) for $m_h/m_e=1$. (c) Difference $\Delta_1(K,K')$ (left) and the second separable scattering minus the third, $ \zeta_2(K) \zeta_2(K')- \zeta_3(K) \zeta_3(K')$,  for $m_h/m_e=5$. (d) Same as (c) for $m_h/m_e=20$. The range of $K$ and $K'$ is $(0,5a^{-1})$. }
   \label{fig3S}
\end{center}
\end{figure}

\subsection{Excitons}

Figure A\ref{fig3S} shows the true angular-averaged scattering $\bar \zeta(K,K')$ and the leading separable scattering $\zeta_1(K)\zeta_1(K')$ for excitons as well as their difference $\Delta_1(K,K')$. As for fermionic-atom dimers, the leading separable scattering already  is in good agreement with the full scattering.\

To improve the result, we note from Fig.~A\ref{fig3S} that $\Delta_1(K,K')$ contains a single peak for $m_h/m_e\ge 16$ (see Fig.~A\ref{fig3S}(d)), a peak plus a dip for $3<m_h/m_e<16$ (see Fig.~A\ref{fig3S}(c)), and a single dip for $m_h/m_e\le 3$ (see Fig.~A\ref{fig3S}(b)). Similar to fermionic-atom dimers, $\Delta_1(K,K')$ can be well simulated by the sum of two separable scatterings, $\zeta_2(K)\zeta_2(K')\pm \zeta_3(K)\zeta_3(K')$, where the $+$ and $-$ signs are used for $m_h/m_e\ge 16$ and $m_h/m_e<16$, respectively. The functions $\zeta_2(K)$ and $\zeta_3(K)$ have the same form as those for fermionic-atom dimers. \

--- For $m_h/m_e\ge 16$, the main peak of $\Delta_1(K,K')$ is simulated by $\zeta_2(K)$ with $\sigma_2=1 a$, and further improved by adding a third separable scattering described by $\zeta_3(K)$, which, in this case, represents a small shoulder peaked at $K_m=K_p/\eta$ with $\eta=0.76+0.1\ln(m_h/m_e)$. The shoulder position shifts as the fermion mass ratio changes. This shifting is also reflected in the minimum position of the exciton-exciton scattering shown in Fig.~\ref{Fig4}. The spread of $\zeta_3(K)$ is determined by $\sigma_3$, which is equal to $1.39a$ for $\ln(m_h/m_e)< 4.2$ and then increases linearly as $\ln(m_h/m_e)$ with a slope $0.16a$.\

--- For $3<m_h/m_e<16$, the peak is followed by a dip. The peak is described by $\zeta_2(K)$ with $\sigma_2=1.3a+1.5 m_e/m_h$ and the dip by $\zeta_3(K)$ with $\sigma_3=1a$, the value being gradually reduced to $0.87a$ as $m_h/m_e$ approaches 3. The extremum points $K_p$ and $K_m$ are chosen to match the corresponding peak/dip positions of $\Delta_1(K,K')$.\

--- For $m_h/m_e \le 3$, only the dip structure remains, and the third separable scattering alone is sufficient to simulate $\Delta_1(K,K')$.  The spread of the dip is best fit by $\sigma_3=0.87a$.


\begin{thebibliography}{99}

\bibitem{Keldysh1968} L. V. Keldysh and A. N. Kozlov, Sov. Phys. JETP {\bf27}, 521 (1968).
\bibitem{HaugPRB1975} H. Haug and E. Hanamura, Phys. Rev. B {\bf11}, 3317 (1975).
\bibitem{ShumwayPRB2001} J. Shumway and D. M. Ceperley, Phys. Rev. B {\bf63}, 165209 (2001); Solid State Comm. {\bf134}, 19 (2005).
\bibitem{Pieri2000} P. Pieri and G. C. Strinati, Phys. Rev. B {\bf61}, 15370 (2000).
\bibitem{Petrov2004}D. S. Petrov, C. Salomon, and G. V. Shlyapnikov, Phys. Rev. Lett. {\bf93}, 090404 (2003); Phys. Rev. A {\bf 71}, 012708 (2005).
\bibitem{Brodsky2006} I. V. Brodsky, M. Yu. Kagan, A. V. Klaptsov, R. Combescot, and X. Leyronas, Phys. Rev. A {\bf73}, 032724 (2006).
\bibitem{Birse2011} M. C. Birse, B. Krippa, and N. R. Walet, Phys. Rev. A {\bf83}, 023621 (2011).
\bibitem{Alzetto2013} F. Alzetto, R. Combescot, and X. Leyronas, Phys. Rev. A {\bf87}, 022704 (2013).


\bibitem{PlatzmannPRB1994} P. M. Platzmann and A. P. Mill, Jr., Phys. Rev. B {\bf49}, 454 (1994).
\bibitem{Ivanov2001} I. A. Ivanov, J. Mitroy, and K. Varga, Phys. Rev. Lett. {\bf87}, 063201 (2001); Phys. Rev. A {\bf65}, 022704 (2002).
\bibitem{AdhikariPL2002} S. K. Adhikari, Phys. Lett. A {\bf294}, 308 (2002).
\bibitem{Chakraborty2004} S. Chakraborty, A. Basu, and A. Ghosh, Nucl. Instrum. Meth. B {\bf221}, 112 (2004).
\bibitem{Daily2015} K. M. Daily, J. von Stecher, and C. H. Greene, Phys. Rev. A {\bf91}, 012512 (2015).

\bibitem{Cassidy2005} D. B. Cassidy et. al., Phys. Rev. Lett. {\bf95}, 195006 (2005).

\bibitem{Avetissian2014} H. K. Avetissian, A. K. Avetissian, and G. F. Mkrtchian, Phys. Rev. Lett. {\bf 113}, 023904 (2014).

\bibitem{Wang2014} Y.-H. Wang, B. M. Anderson, and C. W. Clark, Phys. Rev. A {\bf89}, 043624 (2014).
\bibitem{Cline} J. M. Cline, Z. Liu, G. D. Moore, and W. Xue, Phys. Rev. D {\bf 89}, 043514 (2014).
\bibitem{Berezhiani} L. Berezhiani and J. Khoury, Phys. Rev. D {\bf 92}, 103510 (2015).

\bibitem{MoniqPhysreport} M. Combescot, O. Betbeder-Matibet, and F. Dubin, Physics Reports {\bf 463}, 215 (2008).
\bibitem{book} M. Combescot and S.-Y. Shiau, \textit{Excitons and Cooper Pairs}, Oxford University Press, Oxford (2015).

\bibitem{Moniq2015PRA} M. Combescot, S.-Y. Shiau, and Y.-C. Chang, Phys. Rev. A {\bf93}, 013624 (2016).

\bibitem{Fetter} A. L. Fetter and J. D. Walecka, {\it Quantum Theory of Many-Particle Systems} (McGraw-Hill, New York 1971).
\bibitem{BCS} L. N. Cooper, Phys. Rev. {\bf104}, 1189 (1956).
\bibitem{rich}  R. W. Richardson, Phys. Lett. {\bf3}, 277 (1963); R. W.  Richardson and N.  Sherman, Nucl. Phys. {\bf52}, 221 (1964); R. W. Richardson, J. Math. Phys. {\bf9}, 1327 (1968).




\bibitem{pieri2003PRL} P. Pieri and G. C. Strinati, Phys. Rev. Lett. {\bf91}, 030401 (2003).

\bibitem{MeloPRL1993} C. A. R. S\'{a} de Melo, M. Randeria, and J. R. Engelbrecht, Phys. Rev. Lett. {\bf 71}, 3202 (1993).


\bibitem{Haussman1993} R. Haussmann, Z. Phys. B {\bf91}, 291 (1993).

\bibitem{OBM} O. Betbeder-Matibet and M. Combescot, Eur. Phys. J. B {\bf31}, 517 (2003).

\bibitem{SYAnnals} S.-Y. Shiau, M. Combescot, and Y.-C. Chang, arXiv:1301.7266; Annals of Physics {\bf336}, 309 (2013).
\bibitem{SenEPL2006} A. Sen, S. Chakraborty, and A. S. Ghosh, Europhys. Lett. {\bf76}, 582 (2006).

\bibitem{Jamieson}  M. J. Jamieson , A. Dalgarno, and J. N. Yukich, Phys. Rev. A {\bf46}, 6956 (1992); M. J. Jamieson, A. Dalgarno, and M. Kimura, Phys. Rev. A {\bf 51}, 2626 (1994); C. J. Williams and P. S. Julienne, Phys. Rev. A {\bf47}, 1524 (1995); M. J. Jamieson, and A. Dalgarno, J. Phys. B {\bf31}, L219 (1998).

\end{thebibliography}
\end{document}